\documentclass[twocolumn]{aastex631}

\usepackage{xspace}

\newcommand{\tdes}{\texttt{tdescore}\xspace}
\newcommand{\sncosmo}{\texttt{sncosmo}\xspace}
\newcommand{\rubin}{\textit{Vera C. Rubin Observatory}\xspace}
\newcommand{\ultrasat}{\textit{ULTRASAT}\xspace}
\newcommand{\xgboost}{\texttt{XGBoost}\xspace}
\newcommand{\shap}{\texttt{SHAP}\xspace}

\begin{document}

\title{\tdes: An Accurate Photometric Classifier for Tidal Disruption Events}

\author[0000-0003-2434-0387]{Robert Stein}
\affiliation{Division of Physics, Mathematics, and Astronomy, California Institute of Technology, Pasadena, CA 91125, USA}

\author[0000-0003-2242-0244]{Ashish Mahabal}
\affiliation{Division of Physics, Mathematics, and Astronomy, California Institute of Technology, Pasadena, CA 91125, USA}
\affiliation{Center for Data Driven Discovery, California Institute of Technology, Pasadena, CA 91125, USA}

\author[0000-0002-7788-628X]{Simeon Reusch}
\affiliation{Deutsches Elektronen-Synchrotron (DESY), Platanenallee 6, D-15378 Zeuthen, Germany}
\affiliation{Institut f\"ur Physik, Humboldt-Universit\"at zu Berlin, D-12489 Berlin, Germany}

\author[0000-0002-3168-0139]{Matthew Graham}
\affiliation{Division of Physics, Mathematics, and Astronomy, California Institute of Technology, Pasadena, CA 91125, USA}

\author[0000-0002-5619-4938]{Mansi M. Kasliwal}
\affiliation{Division of Physics, Mathematics, and Astronomy, California Institute of Technology, Pasadena, CA 91125, USA}

\author[0000-0001-8594-8666]{Marek Kowalski}
\affiliation{Deutsches Elektronen-Synchrotron (DESY), Platanenallee 6, D-15378 Zeuthen, Germany}
\affiliation{Institut f\"ur Physik, Humboldt-Universit\"at zu Berlin, D-12489 Berlin, Germany}

\author[0000-0003-3703-5154]{Suvi Gezari}
\affiliation{Space Telescope Science Institute, 3700 San Martin Drive, Baltimore, MD 21218, USA}
\affiliation{Department of Physics and Astronomy, Johns Hopkins University, 3400 N. Charles Street, Baltimore, MD 21218, USA}

\author[0000-0002-5698-8703]{Erica Hammerstein}
\affiliation{Department of Astronomy, University of Maryland, College Park, MD 20742, USA}
\affiliation{Astrophysics Science Division, NASA Goddard Space Flight Center, 8800 Greenbelt Road, Greenbelt, MD 20771, USA}
\affiliation{Center for Research and Exploration in Space Science and Technology, NASA/GSFC, Greenbelt, MD 20771, USA}

\author{Szymon J. Nakoneczny}
\affiliation{Division of Physics, Mathematics, and Astronomy, California Institute of Technology, Pasadena, CA 91125, USA}

\author[0000-0002-2555-3192]{Matt Nicholl}
\affiliation{Astrophysics Research Centre, School of Mathematics and Physics, Queens University Belfast, Belfast BT7 1NN, UK}

\author[0000-0003-1546-6615]{Jesper Sollerman}
\affiliation{Department of Astronomy, The Oskar Klein Center, Stockholm University, AlbaNova, 10691 Stockholm, Sweden}

\author[0000-0002-3859-8074]{Sjoert van Velzen}
\affiliation{Leiden Observatory, Leiden University, Postbus 9513, 2300 RA Leiden, The Netherlands}

\author[0000-0001-6747-8509]{Yuhan Yao}
\affiliation{Miller Institute for Basic Research in Science, 468 Donner Lab, Berkeley, CA 94720, USA}
\affiliation{Department of Astronomy, University of California, Berkeley, CA 94720, USA}

\author[0000-0003-2451-5482]{Russ R. Laher}
\affiliation{IPAC, California Institute of Technology, 1200 E. California Blvd, Pasadena, CA 91125, USA}

\author[0000-0001-7648-4142]{Ben Rusholme}
\affiliation{IPAC, California Institute of Technology, 1200 E. California
             Blvd, Pasadena, CA 91125, USA}

\correspondingauthor{Robert Stein}
\email{rdstein@caltech.edu}



\begin{abstract}

Optical surveys have become increasingly adept at identifying candidate Tidal Disruption Events (TDEs) in large numbers, but classifying these generally requires extensive spectroscopic resources. Here we present \tdes, a simple binary photometric classifier that is trained using a systematic census of $\sim$3000 nuclear transients from the Zwicky Transient Facility (ZTF). The sample is highly imbalanced, with TDEs representing $\sim$2\% of the total. \tdes is nonetheless able to reject non-TDEs with 99.6\% accuracy, yielding a sample of probable TDEs with recall of 77.5\% for a precision of 80.2\%.  \tdes is thus substantially better than any available TDE photometric classifier scheme in the literature, with performance not far from spectroscopy as a method for classifying ZTF nuclear transients, despite relying solely on ZTF data and multi-wavelength catalogue cross-matching. In a novel extension, we use `SHapley Additive exPlanations' (\texttt{SHAP}) to provide a human-readable justification for each individual \tdes classification, enabling users to understand and form opinions about the underlying classifier reasoning. \tdes can serve as a model for photometric identification of TDEs with time-domain surveys, such as the upcoming Rubin observatory.

\end{abstract}



\section{Introduction}

Tidal disruption events occur when stars pass too close to supermassive black holes (SMBHs). The tidal force exerted by the SMBH exceeds the self-gravity holding the star together, and the star disintegrates \citep{rees_88}. Much of the resulting stellar debris remains gravitationally bound to the SMBH, and is ultimately accreted onto the black hole.  These TDEs can generate luminous emission across the entire electromagnetic spectrum, from radio to soft gamma-rays, and in recent years all-sky surveys have become increasingly adept at finding the previously-elusive class of transients \citep[see][for a recent review]{gezari_21}. TDEs offer a unique probe of otherwise-quiescent SMBHs residing in galaxies, and can be used to study a variety of areas such as astrophysical jet launching, SMBH demographics and accretion disk formation. 

There are now $\gtrsim$100 TDEs in the literature, the vast majority of which are identified by optical surveys. In particular, the Zwicky Transient Facility at Palomar Observatory \citep[ZTF;][]{ztf_survey, ztf_obs}  conducts an all-sky survey which has detected $\sim$90 TDEs since 2018 \citep[see e.g][]{ztf_tde_1, final_season,  yao_23}. With this large sample, we now know that at least some TDEs emit quasi-thermal optical flares with high apparent temperature that rise on a timescale of weeks, and fade more slowly over a timescale of months with little apparent temperature evolution \citep{gezari_21}. These optical TDEs appear to have a marked preference for `green-valley' galaxies \cite[see e.g][]{arcarvi_14,french_16,graur_18,hammerstein_21}. 

Despite a nominal survey depth of 20.5 mag \citep{ztf_science}, the ZTF TDE program remains incomplete below a magnitude of $\approx$19.1 mag due to limited spectroscopic resources \citep{yao_23}. This spectroscopic bottleneck will become even more severe with upcoming instruments and observatories such as the \rubin \citep{lsst}, and \ultrasat \citep{ultrasat}, which are expected to detect thousands of TDEs each year \citep[see e.g][]{lsst_tdes, ultrasat}. 

There is thus increasing need for the development of TDE selection methods which do not rely on expensive spectroscopic follow-up. However, photometric classification of nuclear transients remains in its infancy. Although some effort has been devoted to finding TDEs as part of generic multi-modal transient classifiers \citep[see e.g][]{rapid_19, graham_23}, the only effort in the literature which was specifically tailored to TDEs was \cite{fleet_tde}. 

In this letter, we introduce a novel binary machine-learning photometric classifier, \tdes, trained with the sample of ZTF nuclear transients to identify TDEs. The code itself is already available on GitHub\footnote{\url{https://github.com/robertdstein/tdescore}} and Zenodo \citep{tdescore_v1}, while the corresponding training data will be released in a dedicated future publication \citep{sample}. In Section \ref{sec:sample} we introduce this ZTF Nuclear Sample, and in Section \ref{sec:features} we describe the process of generating high-level `features' from the available data. We then outline the \tdes classifier itself (Section \ref{sec:classifier}), and explore the reasoning behind the corresponding classifications (Section \ref{sec:understanding}). Finally, in Section \ref{sec:conclusion}, we highlight the relevance of \tdes to both existing and future surveys.

\section{The ZTF Nuclear Transient Sample}
\label{sec:sample}

The first photometric optical search for TDEs was conducted by \cite{van_velzen_11} using archival searches of Sloan Digital Sky Survey data \citep{sdss_00}, finding that TDEs can be differentiated from supernovae using light curve evolution. 
Photometric identification of TDEs at Palomar began with the predecessor survey to ZTF, the intermediate Palomar Transient Factory (iPTF) survey \citep{iptf}. A systematic census of nuclear transients in 4800 sq. deg. of iPTF data was used to develop simple algorithmic cuts yielding candidate TDEs with a precision of 20\%, which was sufficiently high to serve as a model for spectroscopic surveys \citep{hung_18}. For the ZTF survey, looser cuts were paired with light curve analysis for the nuclear transient filter \citep{ned_stark}, which has been used to identify dozens of TDEs over the course of the survey \citep{ztf_tde_1,final_season, yao_23}. The filter was implemented in AMPEL, a realtime data analysis framework and ZTF alert broker \citep{ampel}. The nuclear transient filter itself is an open-source python script\footnote{\url{https://github.com/AmpelAstro/Ampel-nuclear/blob/main/ampel/nuclear/t0/NuclearFilter.py}}, which broadly selects candidates based on:

\begin{itemize}
    \item estimated `nuclearity' of the flux-weighted ZTF transient position using proximity to sources detected by the deeper Pan-STARRS1 (PS1) survey \citep{panstarrs} 
    \item probability of detection being `real' based on machine-learning \texttt{RealBogus}/\texttt{DeepRealBogus} classification of images \citep{ztf_rb,ztf_drb}, and algorithmic cuts on image detection parameters
    \item rejection of stellar sources via the machine-learning \texttt{sgscore} classification \citep{sgscore} of underlying PS1 sources \citep{panstarrs}, measured parallax in GAIA DR2 \citep{gaia_dr2}, and cuts on bright hosts (m $<$ 12)
    \item rejection of galactic sources by requiring galactic latitude $|b| > 5$
    \item rejection of moving objects by requiring multiple time-separated detections of a source.
\end{itemize}

These cuts are designed to be loose and inclusive, prioritising recall over precision. As part of the ongoing ZTF TDE program, additional light curve analysis and ranking is performed to highlight potential TDE candidates \citep{ztf_tde_1}, which are then vetted by humans and assigned additional follow-up observations for classification. In many cases, a spectrum is required to resolve ambiguity. With \tdes, we aim to develop an alternative to this resource-intensive process using a machine-learning approach.

The nuclear transient filter has been iteratively modified over the course of the survey, to improve the false-positive or false-negative rate. To develop \tdes, we start with the latest version of the filter, which was developed and applied to all archival ZTF alert data, yielding a uniform sample of 11699 nuclear transients discovered in ZTF-I, from 2018 April 1 to 2020 September 30, and in ZTF-II from 2020 October 1 to 2022 April 30 \citep{sample}. 

We extract any available classifications for these transients from the \texttt{ZTF Fritz Marshal}\footnote{\url{https://fritz.science/}}  \citep{skyportal, skyportal_2}, and the predecessor \texttt{ZTF GROWTH Marshal} \citep{growth_marshal}. In general these are accumulated human-assigned classifications which can be based on spectra (including public ones taken from e.g the Transient Name Server, and host spectroscopy from the Sloan Digital Sky Survey \citep{sdss_00}), light curve evaluation or other contextual information. We verify each of these human classifications (see Appendix Section \ref{sec:classifications} for details), and recover 5264 classified sources, of which 86 are classified as TDEs. This includes 30 sources from ZTF-I presented in \cite{ztf_tde_1} and \cite{final_season}, 17 additional bright ZTF-II TDEs from \cite{yao_23}, as well as 39 additional faint or recent TDEs from ZTF-II which have not yet been published.

\section{Feature Extraction}
\label{sec:features}

\subsection{Light Curve Analysis}

To develop a flexible framework which could be easily generalised to other surveys, we use a Gaussian Process to convert the extensive photometry from ZTF into more survey-independent high-level physical features such as peak magnitude and fade rate. We specifically design a multi-step fitting procedure tailored to the known characteristics of TDEs, namely that they are blue, long-lived transients with little apparent colour evolution. Beyond this, the fitting procedure is agnostic about any underlying physical model for TDE emission, and can therefore capture the full diversity of TDE optical emission, including observed TDE outlier behaviour such as multiple peaks or long plateaus.

We use the alert photometry provided directly by ZTF as the basis of the analysis. No K-correction is applied to the data, but we do correct for galactic extinction using results from \cite{schlafly_11} and the extinction law from \cite{fitzpatrick_99}. We perform a series of cuts similar to those in the nuclear filter, to remove detections which are not well subtracted, returning a subset of `clean' photometry for each source. We specifically require a FWHM $<$ 5'', no bad pixels, a Real/Bogus score $>$ 0.3, a pixel distance to host $<$ 1'', and a difference image depth of at least 19.0 mag to reject images taken under poor conditions. Though ZTF provides some (sporadic) $i$-band coverage as part of the partnership surveys, we only consider the $g$-band and $r$-band data, which are primarily provided in a uniform 2-day cadence by the ZTF MSIP public survey.

Our dataset is three dimensional (detections have a flux, wavelength, and time), so we cannot directly apply a simple univariate Gaussian Process. While multi-variate Gaussian Processes have been applied for astronomical datasets, they require a customised covariance matrix to balance variation between bands against variation in time (see e.g \citealt{gp_astro_23} for a recent review). Moreover, multivariate Gaussian Processes have more associated uncertainty in cases such as here where the bands are not sampled uniformly.

Instead, we simplify the problem, and fit flux in one band as a function of time with a univariate Gaussian Process. Given that TDEs are generally blue, we first fit the $g$-band data with a univariate Gaussian process model implemented in \texttt{scikit-learn} \citep{scikit-learn}, using a `Radial Basis Function' (RBF) kernel restricted to timescales of 50d-500d, and an additional white noise component equal to at least 0.1 mag to account for systematic uncertainty and prevent overfitting. After obtaining a model for the $g$-band data, we then perform a least-squares minimisation fit of the $r$-band data to this g-band light curve model, under the assumption that the data follows a linear colour evolution of the form:
\begin{equation}
    m_{r} (t) = m_{g} (t) + C_{0} + (C_{1} \times t)
\label{eq:color}
\end{equation}
where C$_{0}$ and C$_{1}$ are fit parameters derived for each source and \textit{t} is the observer time in days. After obtaining these coefficients, we estimate the $g$-band magnitude of the source for each $r$-band detection.

\begin{figure}
    \centering
    \includegraphics[width=0.45\textwidth]{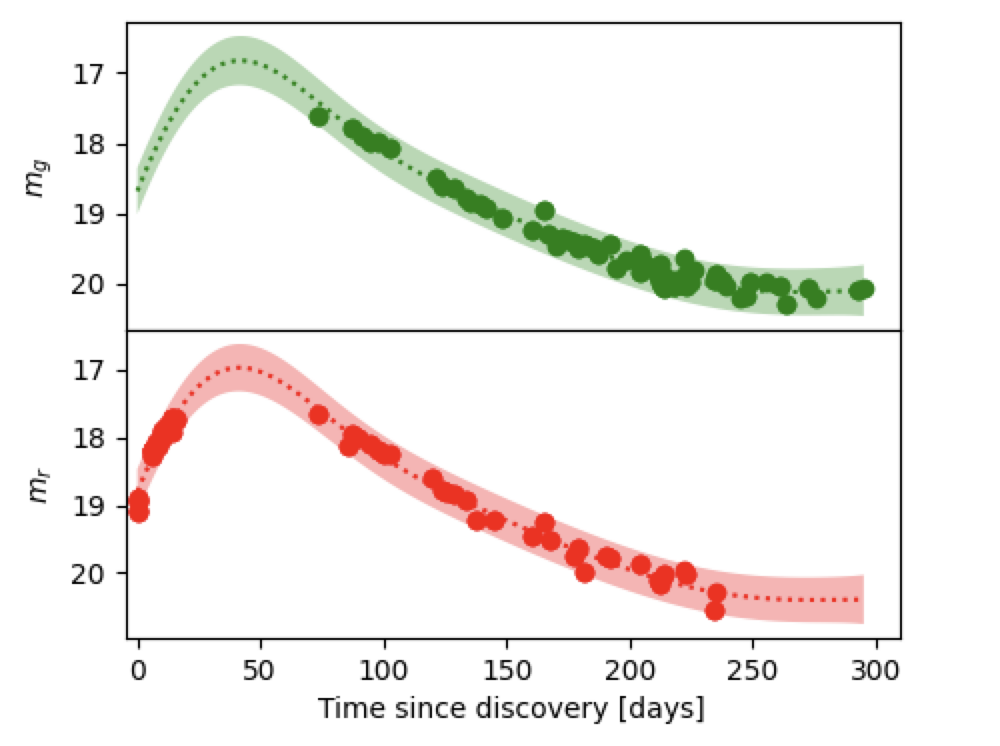}
    \caption{An example of the light curve fitting procedure on a real TDE, ZTF20achpcvt/AT2020vwl \citep{2020vwl_discovery, 2020vwl_classification}, for which limited data was available at peak. Using the two step-fit, the approximate $g$-band peak time, and the colour at peak, can be inferred for use in classification. AT2020vwl is relatively red with $(g-r) \approx0$, but bluer TDEs with $(g-r)>0$ are detected more frequently in $g$-band. Nonetheless, the fitting procedure still works well for this TDE.}
    \label{fig:gp_plot}
\end{figure}

We then fit the combined ($g$-band and converted $r$-band) light curve with the same univariate Gaussian process procedure. This provides our final model for each source light curve. An example of this fitting is shown in Figure \ref{fig:gp_plot}, for a real TDE with sparse early data where the joint fit is required to constrain the $g$-band rise and fade.

With these light curve fits, we can extract high-level parameters for each source. We specifically extract:

\begin{itemize}
    \item the peak magnitude in $g$-band.
	\item the time of peak in $g$-band.
	\item the colour at $g$-band peak (MJD).
	\item the colour change rate (C$_{1}$).
	\item fade time (defined as the time in for the $g$-band light curve to return from peak to 0.5 mag below peak).
	\item the RBF length scale from the Gaussian process fit.
	\item the RBF amplitude  from the Gaussian process fit.
	\item the Gaussian process `score', which quantifies how well the model describes the data.
    \item To fully capture the multiple peaks which can be exhibited by many AGN and some transients, we count the number of inflection points in the light curve fit which occur pre-peak and separately count the post-peak inflection points. 
    \item the mean detection cadence (total number of detections divided by time in days between first and last detection).
\end{itemize}

Entirely independently of the above procedure, we also try to fit the light curves with SALT2 supernova Type Ia models \citep{salt2} using \sncosmo \citep{sncosmo}, and retrieve the underlying c/$x_{1}$ parameters, as well as the $\chi^{2}$, to serve as a proxy for the `Ia'-ness of the light curve. 

When run on a standard MacBook Pro without any parallelisation, the Gaussian process analysis requires $\sim$3s per transient on average. The time varies somewhat between individual transients, with more lightcurve detections leading to longer process times. \sncosmo is faster, requiring $\sim$1s on average per source. The lightcurve analysis procedure is thus fast enough to scale to deeper surveys such as Rubin. For surveys with more than two bands, the model in Equation \ref{eq:color} could be generalised to a thermal model with a temperature and linear temperature evolution.

\subsection{Additional features}

In addition to parameters directly extracted from the ZTF photometry, additional contextual information is extracted for each source. The ZTF alerts themselves \citep{ztf_data, zads} provide the catalogued `\texttt{sgscore}' value for the source host (a binary machine-learning classification score based on morphology to distinguish stars from galaxies \citep{sgscore}). Each individual detection also contains:

\begin{itemize}
    \item \textbf{distpsnr1} - distance of detection to PS1 host in arcseconds, from which we calculate a median.
    \item \textbf{distnr} - pixel distance to nearest source in reference image, from which we calculate a median.
    \item \textbf{sumrat} - The ratio of summed pixels values in a detection to the sum of absolute pixel values, serving as a proxy for yin-yang subtraction artefacts. We calculate a median sumrat for each source.
    \item \textbf{classtar} - Star/Galaxy classification score from \texttt{SourceExtractor} \citep{sextractor}.
    \item \textbf{isdiffpos} - boolean value for whether the detection is positive or negative, from which we calculate an overall fraction of positive detections. 
\end{itemize}

We also cross-match the sources to their underlying PS1 hosts \citep{panstarrs}, yielding $g-r$, $r-i$, $i-z$ and $z-y$ host colours. By construction, all sources will be close to a source with at least one PS1 detection. We also cross-match to MIR host colours (W1$-$W2, W3$-$W4) from WISE \citep{wise}, and to underlying W1 variability using WISE+NEOWISE \citep{neowise}, similar to \cite{yao_23}.
We also crossmatch to the \texttt{Milliquas} catalogue to known radio/X-ray-selected AGN \citep{milliquas}, yielding a boolean \textbf{has\_milliquas} flag.

\section{\tdes}
\label{sec:classifier}

\subsection{ZTF Nuclear ML Dataset}
\label{sec:trainingsample}

From the nuclear sample, we have 5264 sources with classifications which could in principle be used for analysis. The sample is dominated by the 4218 AGN (80.1\%), but also includes 213 core-collapse supernovae (4.0\%), 708 Type Ia supernovae (13.4\%), 39 variable stars (0.7\%), and 86 TDEs (1.6\%). Additional quality cuts are then applied, to select a sample of nuclear transients with uniformly-derived properties. In particular, we restrict ourselves to sources which passed the light curve fitting described in Section \ref{sec:features}, and had a significantly-measured fade time (i.e were detected at least 0.5 mag below peak). In practise, this requires sources to be detected multiple times in both $g$ and $r$ band, and to have a detection at least 0.5 mag below $g$-band peak. Of the initial 5264 classified nuclear sources, only 3040 pass this additional `fade and colour change' cut. All sources passing this step also have the other relevant light curve parameters such as score, colour at peak etc.

\begin{figure}
	\centering
	\includegraphics[width=0.48\textwidth]{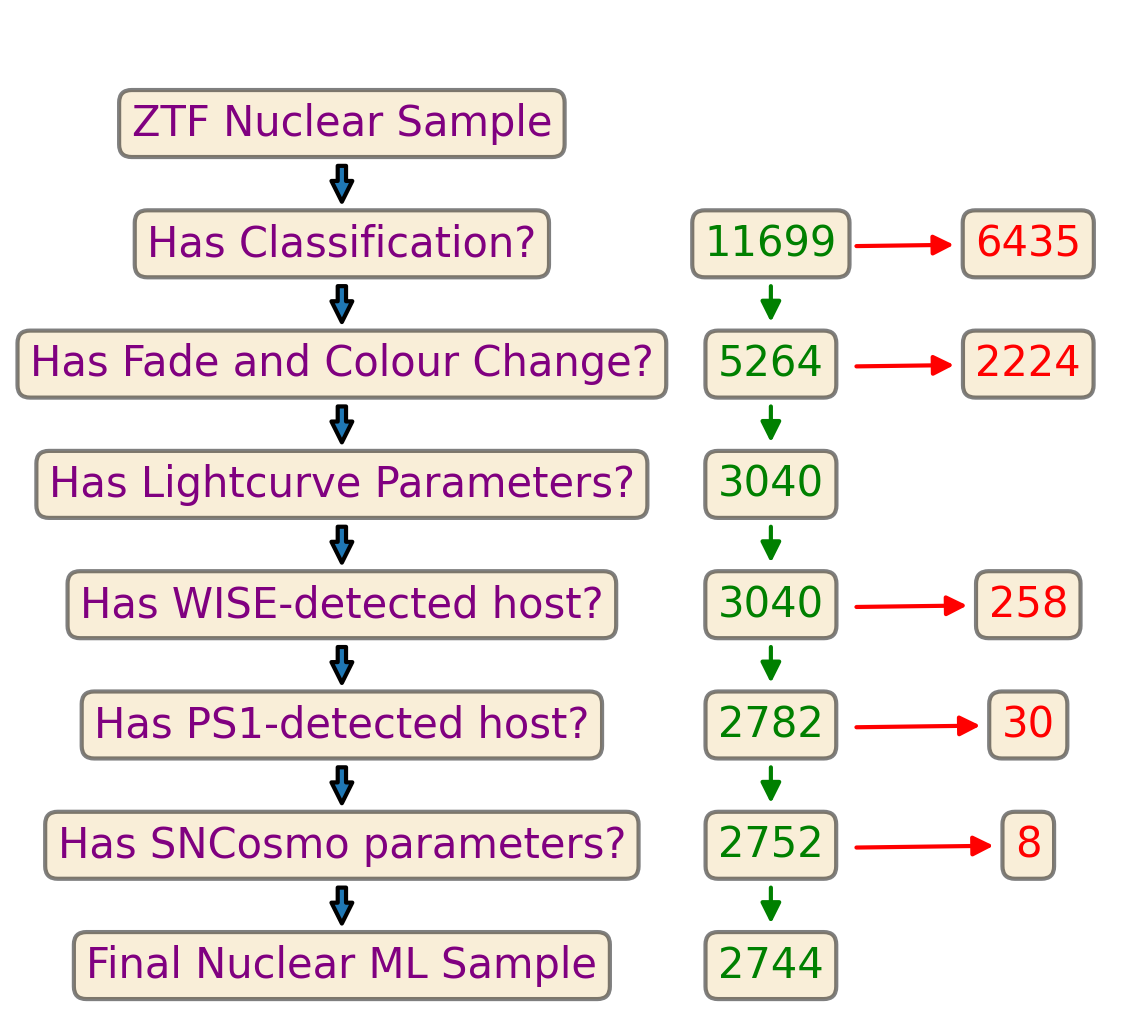}
	\includegraphics[width=0.4\textwidth]{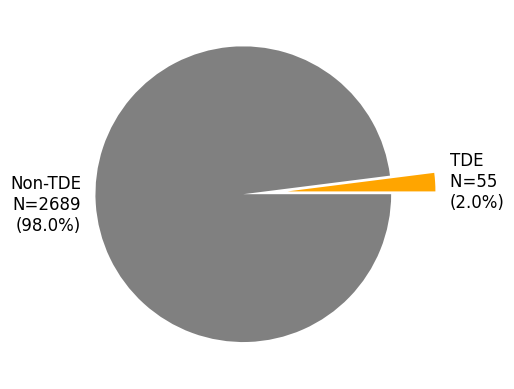}
	\caption{Top: Breakdown of the various cuts applied to the ZTF nuclear sample. Of 11699 ZTF sources, 5264 have a secure classification, 3040 also have a well-measured fade, and 2744 sources pass all cuts. Bottom: Of these 2744 sources used to train \tdes, 55 (2.0\%) are TDEs.}
	\label{fig:pie}
\end{figure}

From these 3040 sources with high-quality light curves, we additionally select those for which all WISE host colours and PS1 host colours were available, and for which \sncosmo successfully ran. Overall, half of the AGN (2153) and CCSNe (106) pass all cuts, along with 60\% of SN Ia (427) and 64\% of TDEs (55). However, only $\sim$8\% variable stars (3) pass, due primarily to their erratic light curves. This ultimately leaves 2744 sources in our final `nuclear ML sample', of which 55 are TDEs and the remaining 2689 are non-TDEs. The share of TDEs thus increases slightly from 1.6\% of classified sources to 2.0\% of the `nuclear ML sample'. These steps are illustrated in Figure \ref{fig:pie}. 

\subsection{Training and Testing Sets}

Given the small number of  TDEs (55) in the dataset, it would not be possible to measure classier performance with reasonable accuracy using a simple division into separate training and testing sets. Even if 20\% of the sources were reserved for testing, this would corresponds to just $\sim$11 TDEs, with consequently high uncertainty for metrics such as recall. Furthermore, given the small number of TDEs, the performance of a classifier on the test sample will be strongly influenced by the randomly-varying composition of the sample. If `atypical TDEs' were randomly to be allocated to the training set, classifier performance would be much better than if they were allocated to the test set.

Instead, to maximise the number of TDEs available for training, and to minimise stochasticity, we employ the `leave one out' k-fold stratified cross-validation to create testing and training sets \citep[see e.g][]{ml_textbook}. We randomly divide our sample into 55 different equally-sized groups, each containing one TDE. The non-TDEs are randomly sorted, and then allocated evenly to one of these groups. As 2689 is not exactly divisible by 55, some groups have 48 non-TDEs, while others have 49 non-TDEs. We select one group to be our test dataset, and use the remaining 54 groups as a training set. After training, we can derive performance metrics on the test dataset. 

We can then repeat the process on a second group, again using the other 54 groups as a training set. This process is repeated  for every single group in the dataset, meaning that 55 different classifiers are trained, with each source being tested once and used for training 55 times. To further reduce the variance in metrics, we repeat the process 10 times, each with a different random sorting of the data. By using the average performance of classifiers across groups and iterations, we can obtain more robust estimates of performance, and be certain that any outlier sources are fairly represented.

\subsection{Dataset Augmentation}
\label{sec:dataaugment}

Given the severe class imbalance in nuclear transients, where TDEs represent a tiny minority ($\sim$2\% of the total), any classifier which simply rejected all candidate TDEs would already have an accuracy of $\sim$98\%. To mitigate this effect, we employ Synthetic Minority Oversampling TEchnique (SMOTE) to generate a balanced training set \citep{smote_concept}. With SMOTE, for each of the k-fold training sets, we randomly select pairs of TDE, and generate new pseudo-TDEs with properties lying a random distance between the two real TDEs. This process is repeated until the training set contains as many TDEs as non-TDEs (and is thus composed of 50\% non-TDEs, $\sim$2\% real TDEs and $\sim$48\% pseudo-TDEs). Once trained on a fold, a classifer can then be tested on the test data, which contains only non-TDEs and real TDEs, to assess its performance. The process of generating pseudo-TDEs via SMOTE is repeated from scratch for each k-fold permutation on the train set, excluding the sources in the test set, so there is no contamination from test data in the training sample.

\subsection{Classifier Architecture and Performance}

With the balanced training sets built in Sections \ref{sec:trainingsample}-\ref{sec:dataaugment}, we can train the \tdes classifier. \tdes is built with the \xgboost algorithm \citep{xgboost}, which employs a gradient-boosted decision tree architecture to build a classifier. For \tdes, we use the \texttt{python} implementation with 27 features. Given the risk of overfitting on our relatively small dataset, and the lack of an independent validation set to measure performance, we generally do not modify the default settings in \xgboost \footnote{For a full explanation of available settings, see: \\\url{https://xgboost.readthedocs.io/en/stable/python/python_api.html\#xgboost.XGBClassifier}}. We use 100 estimators, and to mitigate overtraining, we further adopt a subsampling rate of 70\% for \xgboost to employ in each iteration of the boosting procedure. During training, we use the area under the precision-recall curve as the optimisation metric. Use of this metric ensures that both false positives and false negatives are minimised. The augmentation, training and testing is rapid, requiring approximately 5s for a single iteration on a typical MacBook Pro. 

\begin{figure}
\centering
\includegraphics[width=0.49\textwidth]{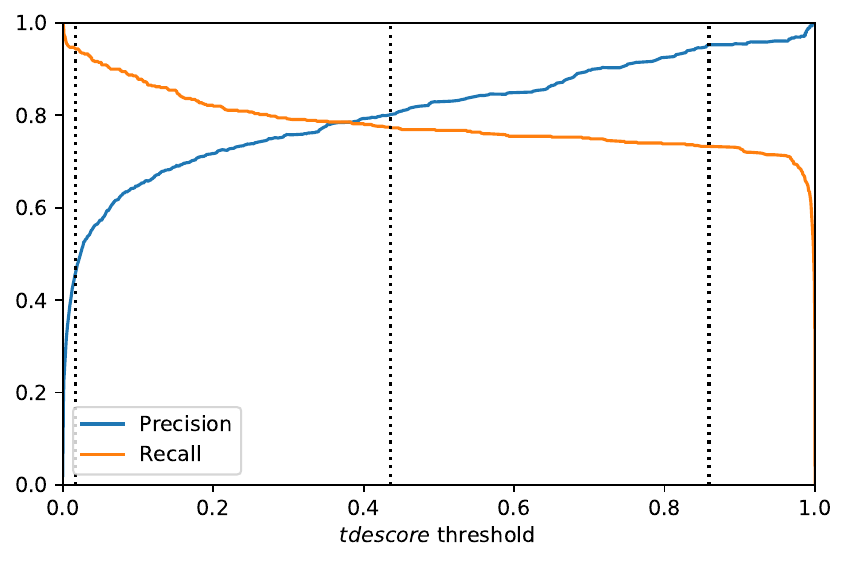} \\
\caption{
Precision and recall as a function of \tdes threshold. The \textit{balanced} threshold (chosen such that precision is at least 80\%) is illustrated by the central vertical line. The \textit{inclusive} ($>$ 95\% recall) and \textit{clean} ($>$ 95\% precision) thresholds are illustrated by the left and right vertical lines, respectively. The corresponding confusion matrices for these three scenarios are shown in Figure \ref{fig:matrix}.}
\label{fig:pr}
\end{figure}

\begin{figure*}
\centering
\includegraphics[width=0.49\textwidth]{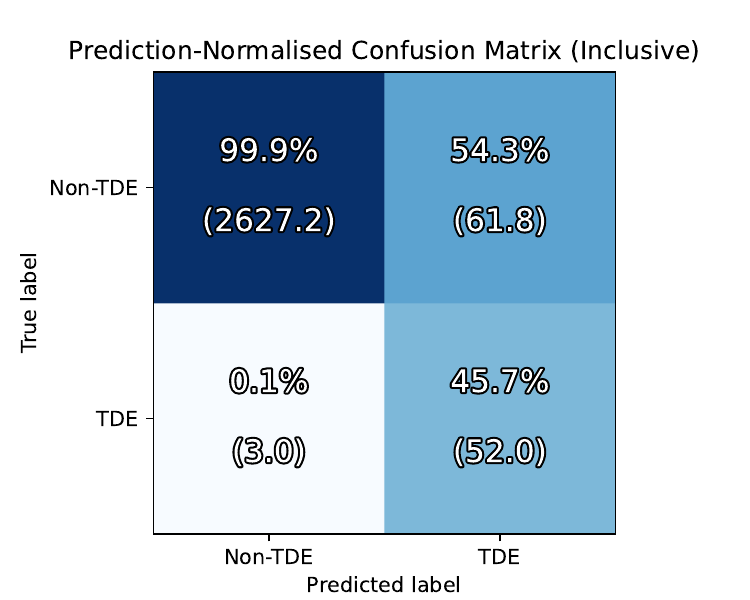}
\includegraphics[width=0.49\textwidth]{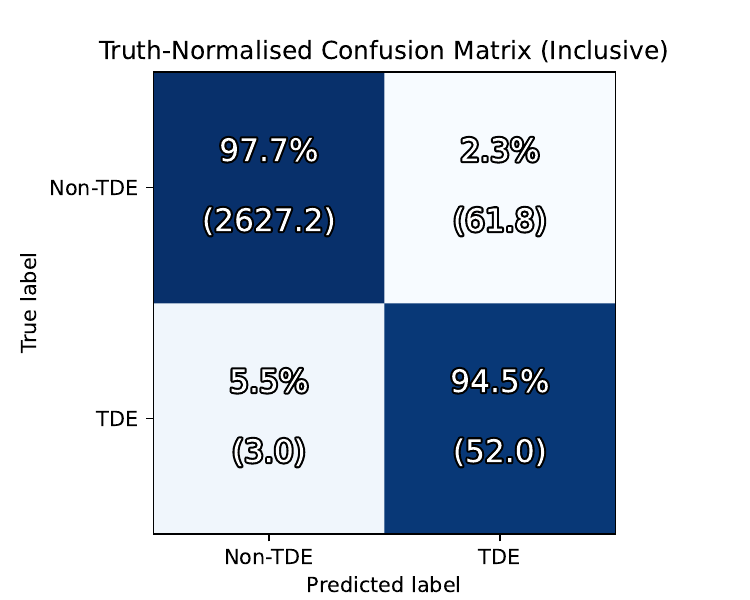}
\includegraphics[width=0.49\textwidth]{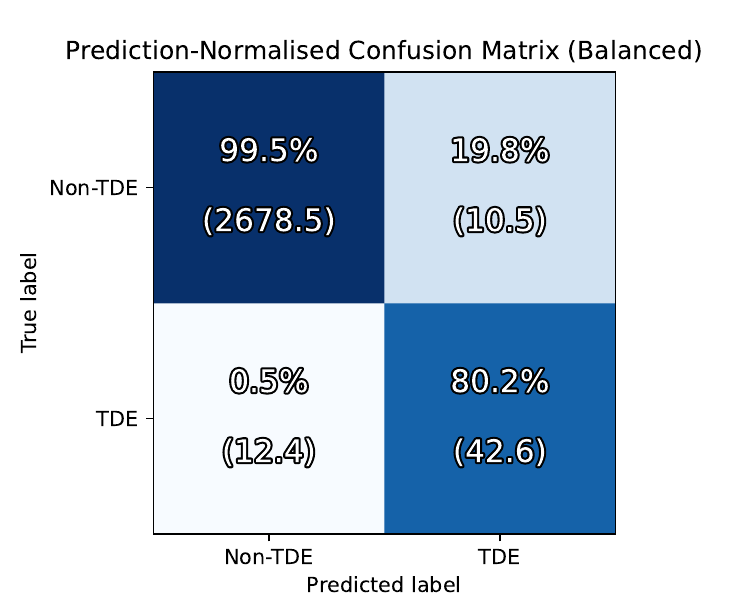}
\includegraphics[width=0.49\textwidth]{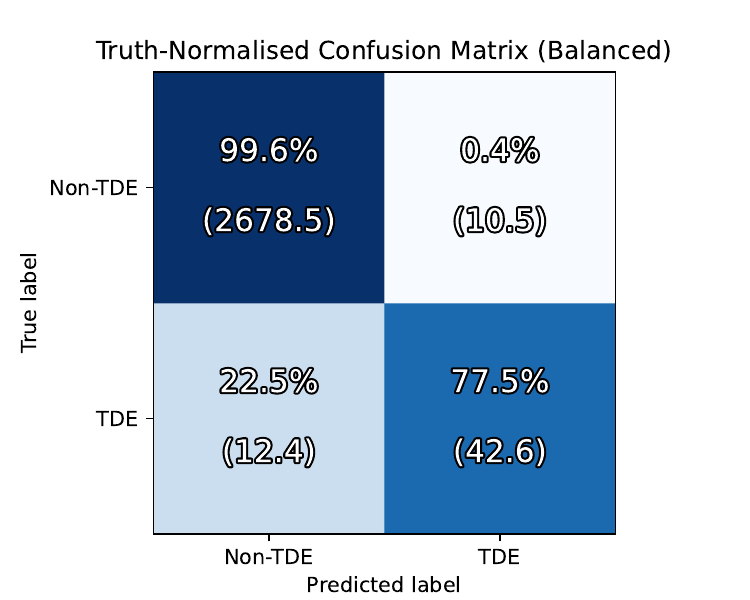}
\includegraphics[width=0.49\textwidth]{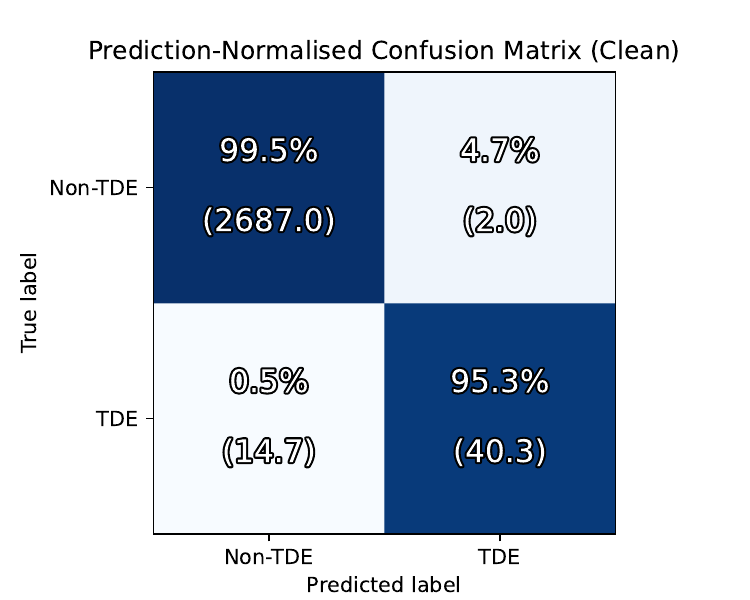}
\includegraphics[width=0.49\textwidth]{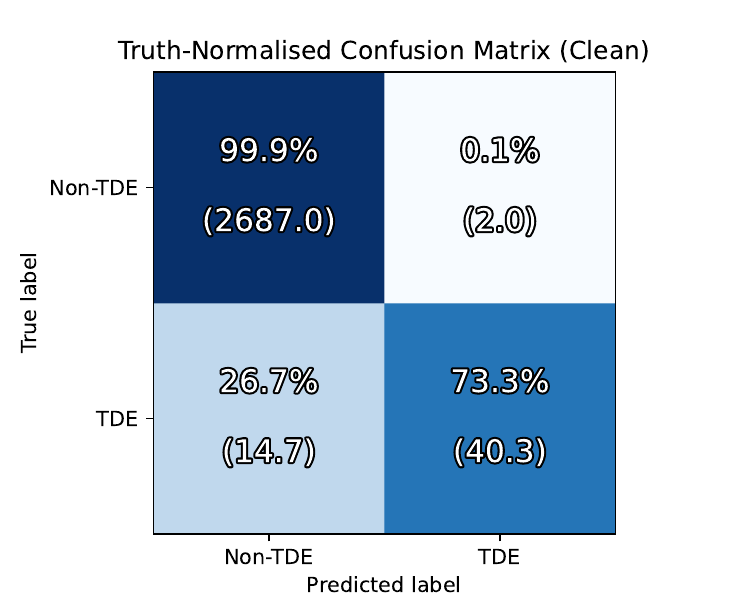}
\caption{
Prediction-normalised confusion matrices (\textbf{left}),  and truth-normalised confusion matrices (\textbf{right}), showing the performance of \tdes on the real data for different thresholds.  The dataset is highly imbalanced, as seen in Figure \ref{fig:pie}. The source shuffling is performed 10 times, yielding averaged performance across the iterations, with the average expected number of sources for each category shown in brackets. 
\textbf{Top:} An inclusive threshold, optimised for recall. At the cost of 5\% loss of TDEs, a sample is produced with a TDE fraction increased from $\sim$2\% to $\sim$46\%. \textbf{Center:} An intermediate threshold, chosen to achieve $>$80\% precision. It achieves relatively high recall (77.5\%).  \textbf{Bottom:} A strict threshold, optimised for precision. $>$70\% of TDEs pass this requirement, yielding a clean sample with $<$5\% contamination rate.}
\label{fig:matrix}
\end{figure*}

Having trained our classifier and applied it to the entire Nuclear ML sample, we then require a threshold score to determine which class each source is assigned. The precision and recall as a function of possible threshold is illustrated in Figure \ref{fig:pr}. As our base case, we adopt a threshold at which $>$ 80\% precision\footnote{`precision' is often called `purity' in astronomical contexts.} is achieved, with the corresponding confusion matrices shown in Figure \ref{fig:matrix}. With this cut, 77.5\% of TDEs are successfully recovered ($\sim$43 TDEs). The classifier efficiently rejects non-TDEs, with 99.6\% being correctly classified, while just 0.4\% are misclassified as TDEs ($\sim$11 non-TDEs). Given the unbalanced sample, this results in 80.2\% of \tdes-selected candidates being real TDEs, with 19.8\% being non-TDEs.

The appropriate threshold for classifiers such as \tdes ultimately depends on the intended scientific application. A high precision sample with lower recall\footnote{`recall' is synonymous with `completeness'.}  may be preferable for rate studies or other population analysis, whereas a high recall might be desired to generate a complete spectroscopically-classified TDE sample where some contamination is acceptable. We consider an alternative stricter threshold, chosen such that at least 95\% of \tdes-selected TDEs would be genuine. Applying this higher threshold produces a very clean sample of probable TDEs, which nonetheless retains a recall of 73.3\% ($\sim$40 TDEs and $\sim$2 non-TDEs). This confirms that nearly three quarters of genuine TDEs are confidently identified, receiving very high classifier scores. We also consider a loose threshold that is nearly complete, chosen such that a recall of at least 95\% is achieved. With this loose cut, only $\sim$5\% of TDEs are lost ($\sim$3 TDEs), but the background is rejected with such efficiency (97.4\%) that the share of TDEs in the sample reaches 45.7\% ($\sim$52 TDEs and $\sim$62 non-TDEs), versus just 2.0\% in the parent training sample. \tdes is thus able to reject most of the background at very little cost to completeness. Further tests of \tdes using subsets of the parameters are detailed in Appendix Section \ref{sec:param_appendix}, which confirm that much of the background can be rejected even before lightcurve information is available.

As a crosscheck, we repeat the \tdes training without using the SMOTE augmentation described in Section \ref{sec:dataaugment}. For the balanced threshold (defined as $>$80\% precision), recall slightly increases from 77.5\% to 79.5\%, but for the clean threshold, recall falls from 73.3\% to 72.0\%. For the inclusive threshold (defined as $>$95\% recall), precision falls substantially from 45.7\% to 29.5\%. Overall, the area under the precision/recall curve decreases from 0.893 to 0.882. The data augmentation step thus provides clear performance improvements for cases prioritising either high recall or precision.



\section{Understanding classifier reasoning}
\label{sec:understanding}


To have confidence in the results of \tdes, it is important to understand whether classifications are based on sound reasoning. The global importance of different features are listed in Table \ref{tab:importance}. In agreement with \cite{fleet_tde}, we find that color at peak is an important discriminator, confirming the well-known property that TDEs are atypically blue relatively to most other transients. However, given the overwhelming dominance of AGN as contaminant nuclear sources, we find that WISE W1$-$W2 colour is by the far the most important feature in identifying TDEs. This is not unexpected, given the ubiquity of WISE colour cuts as a method of selecting AGN \citep{stern_12}. We also find that \sncosmo analysis can be a useful tool, with the resultant $\chi^{2}$ values being useful proxies for both SNIa (with good fits) and AGN (with poor fits).

\begin{table*}[]
\centering
    \begin{tabular}{c|c|c}
    \textbf{Feature} &\textbf{Description}& \textbf{Importance (\%)}\\
    \hline

	w1\_m\_w2 & WISE W1$-$W2 host colour & 32.4 \\
	peak\_color & Colour at g-band peak & 16.0 \\
	has\_milliquas & Has milliquas crossmatch? & 9.3 \\
	color\_grad & Rate of colour change & 7.8 \\
	sncosmo\_chisq & sncosmo $\chi^{2}$ & 5.9 \\
	sncosmo\_c & sncosmo c parameter & 4.6 \\
	fade & Fade from G.P. & 3.8 \\
	det\_cadence & Mean detection candence & 2.4 \\
	pre\_inflection & Number of pre-peak inflections & 2.3 \\
	distpsnr1 & Distance to PS1 host & 1.8 \\
	length\_scale & Length scale from G.P. & 1.6 \\
	y\_scale & Y Scale from G.P. & 1.6 \\
	sncosmo\_x1 & sncosmo X1 parameter & 1.6 \\
	w3\_m\_w4 & WISE W3$-$W4 host colour & 1.2 \\
	post\_inflection & Number of post-peak inflections & 1.0 \\
	g-r\_MeanPSFMag & PS1 host $g-r$ colour & 0.9 \\
	sumrat & `Sum ratio' & 0.9 \\
	score & Score from G.P & 0.8 \\
	classtar & \texttt{SourceExtractor} variable & 0.8 \\
	positive\_fraction & Fraction of positive detections & 0.6 \\
	w1\_chi2 & WISE W1 $\chi^{2}$ & 0.6 \\
	distnr & Pixel distance to nearest source & 0.6 \\
	z-y\_MeanPSFMag & PS1 host $z-y$ colour & 0.4 \\
	sncosmo\_chi2pdof & sncosmo $\chi^{2}$ per d.o.f & 0.4 \\
	i-z\_MeanPSFMag & PS1 host $i-z$ colour & 0.2 \\
	r-i\_MeanPSFMag & PS1 host $r-i$ colour & 0.1 \\
	sgscore1 & Star/Galaxy Score for PS1 host & 0.1 \\
\end{tabular}
\caption{Relative importance of all 27 features in \tdes, calculated by \xgboost \citep{xgboost} using the standard averaging of importance across all decision trees in the final model \citep[see e.g][]{ml_textbook}.}
\label{tab:importance}
\end{table*}

\begin{figure}[t]
\centering
\includegraphics[width=0.49\textwidth]{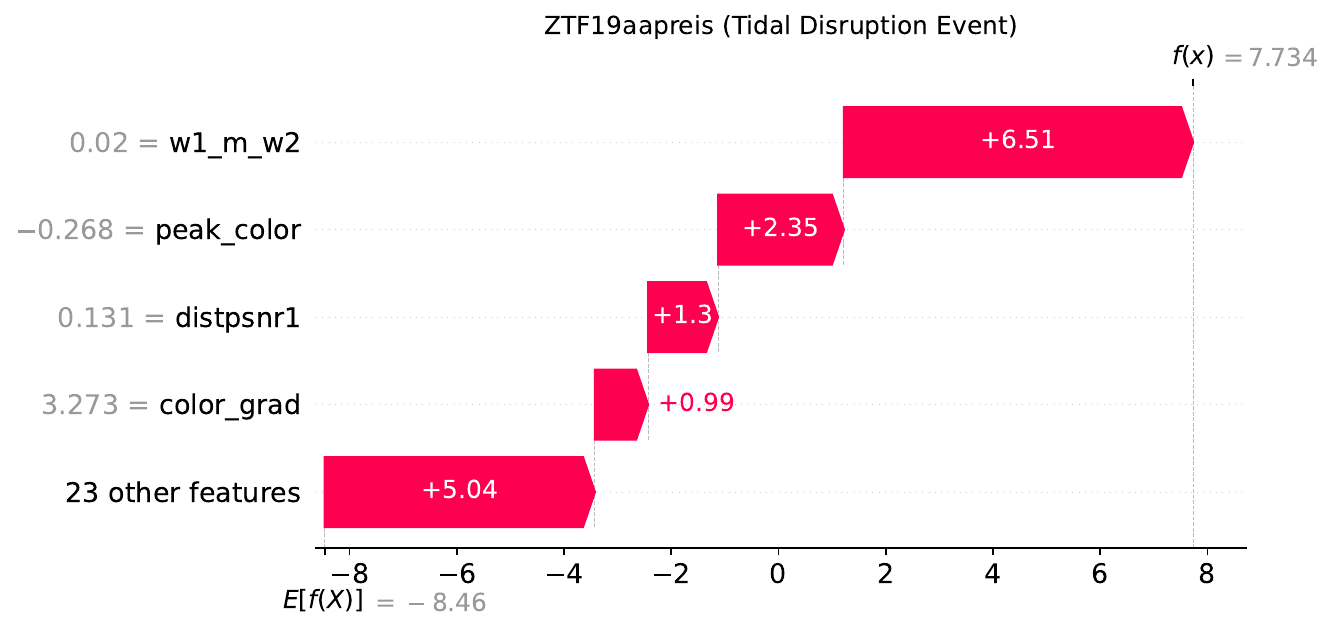}
\includegraphics[width=0.49\textwidth]{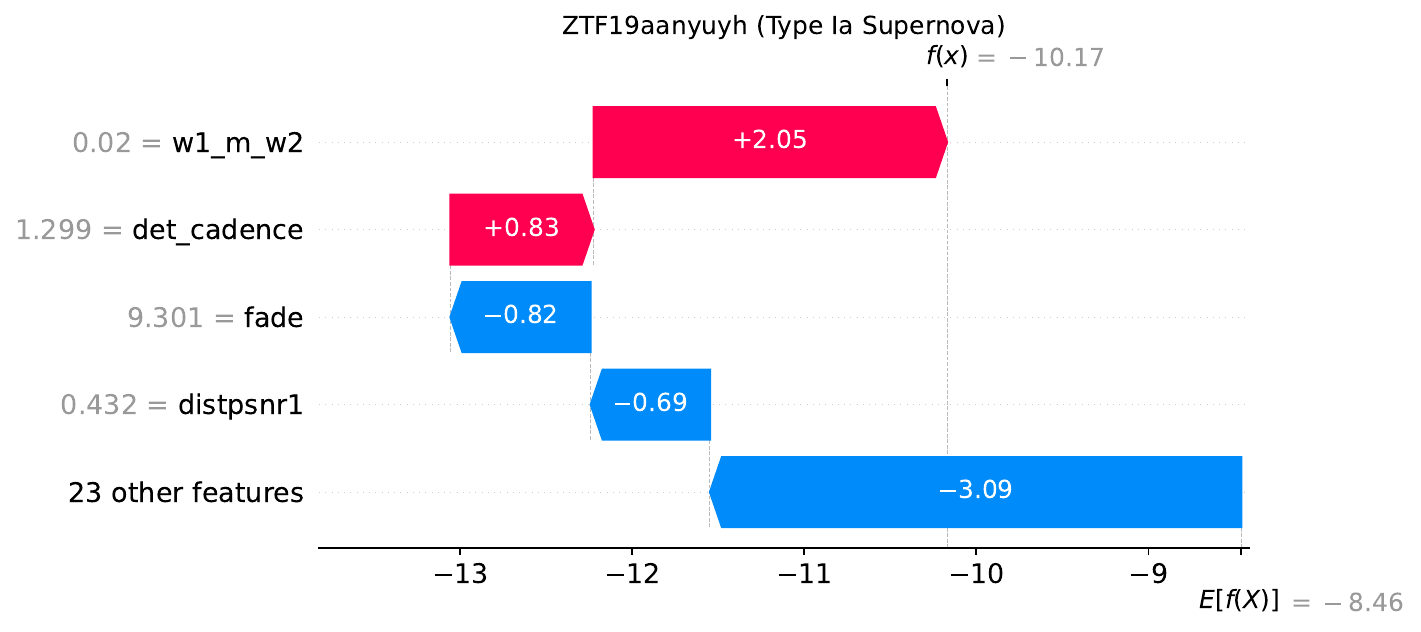}
\caption{`Waterfall plots' produced by SHAP for a TDE (top) and a supernova Type Ia (bottom), demonstrating the thinking behind the \tdes classifications. In both plots, red/right is more TDE-like, while blue/left is less TDE-like. The four most salient features for each source are shown, with the actual value for each parameter given in the left-most column. \textbf{Top:} The TDE (ZTF19aapreis) has WISE colours inconsistent with an AGN host (W1$-$W2 $=0.0$), a blue colour at peak ($g-r=-0.3$), is very nuclear (0.1" offset to PS1), and has very little cooling (0.003 mag per day). All these variables lead to a TDE classification. \textbf{Bottom:} The supernova (ZTF19aanyuyh) also has a WISE colours inconsistent with an AGN host (W1$-$W2 $=0.0$), and a high detection rate (one datapoint per 1.3 days), supporting a possible TDE classification. However, the source also fades very rapidly (9.3 days), and is somewhat offset from its PS1 host (0.4''). In combination, these other variables lead to a non-TDE classification. In both cases, the \tdes use of features closely approximates the reasoning that would be employed by an astronomer.}
\label{fig:explain}
\end{figure}

\tdes also attempts to overcome the `black-box-problem' by incorporating \emph{explainable AI}. We analyse the \tdes classifier using SHapley Additive exPlanations (\shap) python package \citep{shap}. \shap explains the output of ML classifiers for individual objects, by estimating the local importance of each feature for a given source. This means that every individual \tdes classification can readily be understood and sanity-checked by humans. An illustration of \tdes reasoning for classifying a TDE, and a Type Ia supernova, are shown in Figure \ref{fig:explain}. In these cases, \tdes follows a decision-making process very similar to that employed by human scanners in ZTF.

\section{Discussion and Conclusion}
\label{sec:conclusion}

\tdes is a novel photometric classifier developed with the explicit aim of approximating the human scanning employed in ZTF. Our ZTF sample provides the largest homogeneous sample of nuclear transients by far \citep{yao_23}, and thus presently serves as the best template for developing techniques to detect TDEs. \tdes combines well-tested algorithmic cuts to robustly identify nuclear transients, an agnostic light curve analysis technique using Gaussian processes, and a simple binary tree-based classifier using physically-motivated features. 

The sole other dedicated TDE classifier in the literature, \texttt{fleet} \citep{fleet_tde}, is based on an adapted supernova classifier. \cite{fleet_tde} began with a sample of spectroscopically-classified transients from the Transient Name Server, rather than a dedicated sample of nuclear transients as presented here. In other respects \cite{fleet_tde} followed a similar procedure to the one presented here, with an imbalanced sample of transients which are first analysed for light curve and host galaxy properties, augmentation via SMOTE and then performance assessment via k-fold cross-validation.
\texttt{fleet} achieved just $\approx$40\% recall with $\approx$50\% precision for a loose selection, or alternatively $\approx$30\% recall with $\approx$80\% precision for a stricter selection,  in contrast to the $\sim$80\% recall and $\sim$80\% precision in the \tdes balanced case. However, the performance is not directly comparable, because \texttt{fleet} was applied to only 40 days of photometry, rather than the full light curve history employed here. For a TDE such as that in Figure \ref{fig:gp_plot}, 40 days would be insufficient to adequately measure fade or colour evolution. As detailed in the Appendix Section \ref{sec:param_appendix}, the performance of \tdes is closer to \texttt{fleet} if late-time data is ignored.

Looking further ahead, \tdes can serve as a template for obtaining a photometrically-selected sample of TDEs from surveys such as the Legacy Survey of Space and Time (LSST) with the \textit{Vera C. Rubin Observatory} \citep{lsst}. In combination with photometric redshifts, an ML-based approach like \tdes could enable us to perform large-sample TDE demographic studies for the first time without use of any spectroscopic observations. In particular, \cite{lsst_tdes} estimated that LSST should detect $>$3000 TDEs per year, under the assumption of a conservative detection requirement of 2 magnitudes above the median 5-sigma limit. Pushing one magnitude deeper, to match the cuts employed by this work, would increase this number even further. The performance of \tdes suggests such a depth would be plausible using photometric selection, with the slow evolution of TDEs being well-suited to the expected LSST cadence.


The performance of \tdes with real-time ZTF data will be the subject of a future publication. There are many other possible uses of photometrically-selected TDEs, for example to build a much larger sample of probable TDEs to test possible multi-messenger correlations between neutrinos and TDEs \citep[see e.g][]{icrc_19}, for which there is growing evidence \citep{bran, tywin, lancel, jiang_23}. Another use is to quickly identify candidate TDEs amongst transients detected by surveys at other wavelengths, through crossmatching to probable ZTF TDEs found by \tdes. We will use this method to aid searches for dust-obscured TDEs with the Wide-Field Infrared Transient Explorer \citep[WINTER;][]{winter}, a newly-commissioned near-infrared survey telescope at Palomar Observatory. 

Building broader TDE samples is important, because by construction, \tdes will not find TDEs that differ substantially from the existing ZTF TDE sample. In particular, given the importance of the W1$-$W2 colour, \tdes is likely to be heavily biased against finding TDEs in AGN. This is a direct consequence of the parent sample of ZTF TDEs, none of which occur in AGN-like hosts with W1$-$W2 $>$ 0.7 \citep{stern_12}. To find such `AGN$-$TDEs' (or other outliers such as red TDEs or fast TDEs), we would first require a handful of spectroscopically-confirmed ZTF examples. As our understanding of TDE diversity improves, \tdes can be retrained to find a broader selection of TDEs. 

Applying \tdes directly to future optical surveys should be relatively straightforward, because the classifier is trained almost exclusively on light curve features that are generic, and do not encode any specific ZTF survey information. However, there is also substantial scope for improvement in performance. While all ZTF light curves were analysed here in observer frame units, with no correction for redshift, ongoing industrial spectroscopic surveys such as DESI \citep{desi} mean that spectroscopic redshifts will be available systematically for much of the local universe. Even in the LSST/Rubin era, widespread adoption of photometric redshifts would enable intrinsic rest-frame properties such as peak luminosity to be employed for classification. Additionally, TDEs are generally characterised by luminous UV emission, and $u$-band colour is  an excellent discriminator to find TDEs \citep[see e.g. ][]{van_velzen_11}. While no UV observations were used for \tdes, due to a lack of systematic coverage, Rubin will have $u$-band coverage of all transients on a $\sim$weekly cadence. At higher redshifts, much of the TDE rest-frame emission at UV wavelengths will also be detectable with optical LSST filters. There are thus many reasons to be optimistic that future iterations of \tdes will be able to outperform the classifier presented here. 

\begin{acknowledgments}
\section*{Acknowledgements}
We thank Adam Stein, Ludwig Rauch, and Niharika Sravan for fruitful discussions about machine learning classification. 

RS and M.M.K acknowledge support from grants by the National Science Foundation  (AST 2206730) and the David and Lucille Packard Foundation (PI Kasliwal). MN is supported by the European Research Council (ERC) under the European Union’s Horizon 2020 research and innovation programme (grant agreement No.~948381) and by UK Space Agency Grant No.~ST/Y000692/1. EKH acknowledges support by NASA under award number 80GSFC21M0002. SJN is supported by the US National Science Foundation (NSF) through grant AST-2108402

Based on observations obtained with the Samuel Oschin Telescope 48-inch and the 60-inch Telescope at the Palomar Observatory as part of the Zwicky Transient Facility project. ZTF is supported by the National Science Foundation under Grants No. AST-1440341 and AST-2034437 and a collaboration including current partners Caltech, IPAC, the Weizmann Institute of Science, the Oskar Klein Center at Stockholm University, the University of Maryland, Deutsches Elektronen-Synchrotron and Humboldt University, the TANGO Consortium of Taiwan, the University of Wisconsin at Milwaukee, Trinity College Dublin, Lawrence Livermore National Laboratories, IN2P3, University of Warwick, Ruhr University Bochum, Northwestern University and former partners the University of Washington, Los Alamos National Laboratories, and Lawrence Berkeley National Laboratories. Operations are conducted by COO, IPAC, and UW. SED Machine is based upon work supported by the National Science Foundation under Grant No. 1106171. The Gordon and Betty Moore Foundation, through both the Data-Driven Investigator Program and a dedicated grant, provided critical funding for SkyPortal. 
\end{acknowledgments}

%

\facilities{PO:1.2m \citep[ZTF;][]{ztf_survey}, PO:1.5m \citep[SEDm;][]{blagorodnova_18,rigault_19,kim_22}}


\software{\texttt{AMPEL} \citep{ampel}, \texttt{astropy} \citep{astropy}, astroquery \citep{astroquery}, \texttt{numpy} \citep{numpy}, \texttt{pandas} \citep{pandas}, \texttt{scikit-learn} \citep{scikit-learn}, \texttt{scipy} \citep{scipy}, \texttt{SHAP} \citep{shap}, \sncosmo \citep{sncosmo}, \tdes \citep{tdescore_v1}, \xgboost \citep{xgboost}}

\clearpage

\appendix

\section{Classification of ZTF 
Nuclear transients}
\label{sec:classifications}

With the 11699 transients in our sample, we employ a two step process to assign classifications to each source. We extract available classifications for these transients from the \texttt{ZTF Fritz Marshal}  \citep{skyportal, skyportal_2}, and the predecessor \texttt{ZTF GROWTH Marshal} \citep{growth_marshal}. However, these classifications have been assigned by human scanners in a variety of heterogeneous ways. While our sample of TDEs has been thoroughly vetted, we cannot vet the justification and reliability of each individual classification assigned for the $>$5000 non-TDEs in this way. Instead, we independently classify this sample of nuclear sources using objective contextual data, and cross-check to confirm that the classifications agree with the human-assigned labels.

We follow a hierarchical approach for the non-TDEs, and first consider classifications based directly on spectra of transients:

\begin{itemize}
    \item \textbf{TNS Spectra:} Transients which are classified on TNS require an accompanying spectrum. We assume that these TNS classifications are accurate. A total of 953 transients have a TNS classification.
    \item \textbf{ZTF Spectra:} We assume that internal ZTF classifications are reliable if at least one spectrum of the source is available. A total of 328 transients are not classified on TNS but do have an internal ZTF classification and spectrum.
\end{itemize}

We further consider sources which have archival spectroscopy confirming a variable host:

\begin{itemize}
    \item \textbf{SDSS Spectra:} We crossmatch every source to the catalogue of SDSS spectra \citep{sdss_00}. We assume that the classifications are reliable if a human scanner has classified a source as variable, and the source also has a spectroscopic classification confirming it as variable (AGN or stellar) from SDSS. There are 1653 such sources.
\end{itemize}

Finally we consider sources which are very likely to be variables:

\begin{itemize}
    \item \textbf{Milliquas AGN:} We consider a classification to reliable if a source is listed in of the `Million Quasar Catalogue' (Milliquas V8) of known AGN \citep{milliquas}, and has also been classified by a human an AGN. 688 sources meet this criteria.
    \item \textbf{WISE AGN:} We consider sources with very AGN-like WISE colours (W1$-$W2) $>$ 0.8 to be reliably classified \citep{stern_12}. There are 973 such sources classified via WISE colours.
    \item \textbf{Gaia QSOs:} We consider Gaia-DR3 catalogued QSOs \citep{gaia_dr3_extragalactic} as AGN, if this agrees with a human classification. 669 sources meet this criteria.
\end{itemize}

\begin{figure}
\includegraphics[width=0.5\textwidth]{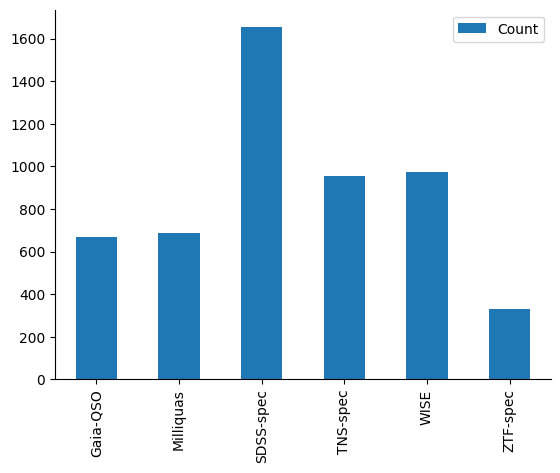}
    \centering
    \caption{Breakdown of the validation method for classifications, as described in Section \ref{sec:classifications}. Each source requires both a human-assigned classification and a second piece of confirmatory evidence to be considered reliably classified.}
    \label{fig:origin}
\end{figure}

After following this procedure, a total of 5264 sources have a verified classification. We omit the remaining ambiguous sources, and treat these 5264 verified objects as our final sample of `classified sources'. The breakdown in classification origin is given in Figure \ref{fig:origin}. 

\section{Performance of \tdes with different parameter subsets}
\label{sec:param_appendix}

We tested the performance of \tdes using subsets of parameters listed in Table \ref{tab:parameter_subset}. For consistency, we measure performance on the same 2744 sources for which all information is available. The parameters were grouped into `Host Only', `Early', `At Peak' and `Full', where full corresponds to the complete parameter set described in Section \ref{sec:features}. The parameter sets were cumulative, with e.g the `Early' set including all `Host' parameters, to provide an estimate of performance over time as available data increased.

\begin{figure}[t]
    \centering
    \includegraphics[width=0.49\textwidth]{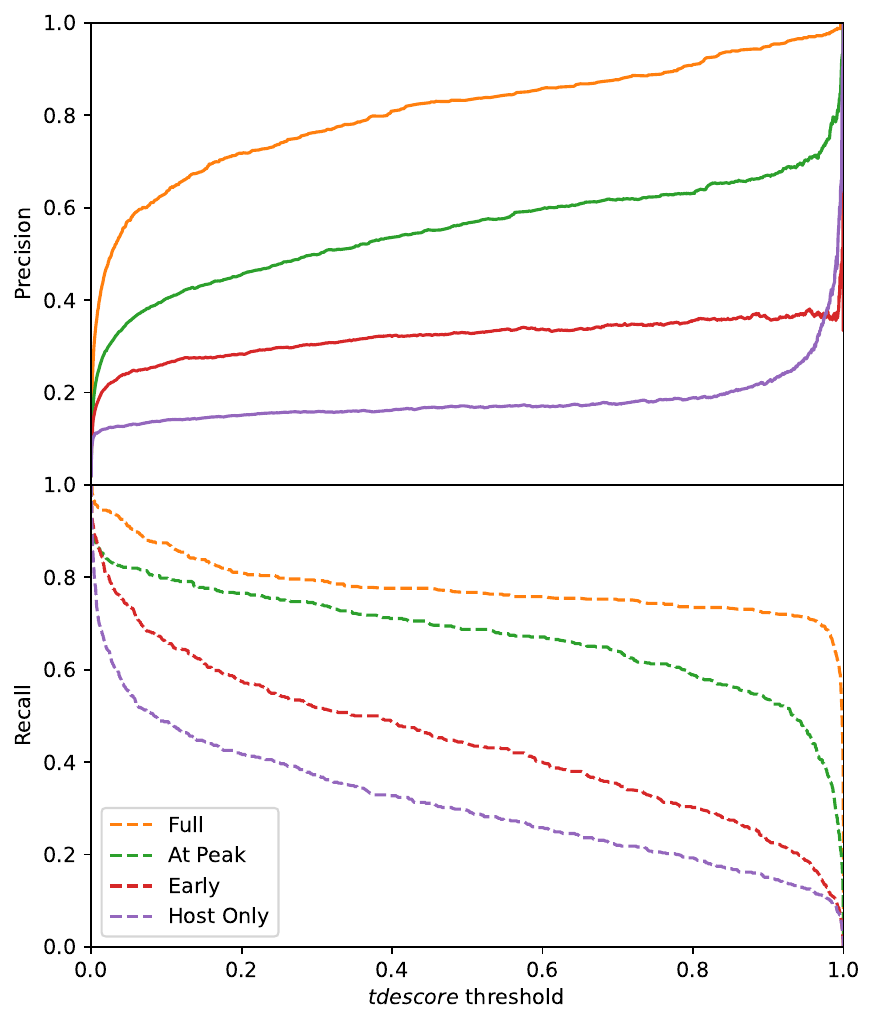}
    \caption{Precision (top, solid) and recall (bottom, dashed) curves for the four parameter sets listed in Table \ref{tab:parameter_subset}, as a function of threshold. Both precision and recall increase substantially as more data is added.}
    \label{fig:pr_comparison}
\end{figure}

We measure the performance of \tdes using each of the four datasets, with the precision/recall area and ROC area listed in Table \ref{tab:parameter_subset}. We also show the full precision and recall of each classifier in Figure \ref{fig:pr_comparison}. As expected, performance improves with increased parameter numbers. However, we note that using only data available after first detections, a recall of $\sim$50\% and precision of $\sim$30\% is achieved. \tdes can thus identify candidate TDEs with a moderate false positive rate at this stage, making triggered follow-up of `infant TDEs' feasible.

\begin{table*}[]
        \begin{tabular}{c|c|c|c|c}
        \textbf{Parameter Set} & \textbf{New Parameters} & \textbf{Total Parameters} & \textbf{ROC Area} & \textbf{Precision/Recall Area} \\
        \hline

	 \textbf{Host Only} & sgscore1 & 9 & 0.91 & 0.20 \\
	 & w1\_m\_w2 &  &  & \\
	 & w3\_m\_w4 &  &  & \\
	 & w1\_chi2 &  &  & \\
	 & has\_milliquas &  &  & \\
	 & g-r\_MeanPSFMag &  &  & \\
	 & r-i\_MeanPSFMag &  &  & \\
	 & i-z\_MeanPSFMag &  &  & \\
	 & z-y\_MeanPSFMag &  &  & \\
	 \hline
	 \textbf{Early} & distpsnr1 & 13 & 0.95 & 0.30 \\
	 & classtar &  &  & \\
	 & sumrat &  &  & \\
	 & distnr &  &  & \\
	 \hline
	 \textbf{At Peak} & peak\_color & 18 & 0.96 & 0.63 \\
	 & pre\_inflection &  &  & \\
	 & positive\_fraction &  &  & \\
	 & det\_cadence &  &  & \\
	 & y\_scale &  &  & \\
	 \hline
	 \textbf{Full} & color\_grad & 27 & 0.99 & 0.89 \\
	 & fade &  &  & \\
	 & length\_scale &  &  & \\
	 & post\_inflection &  &  & \\
	 & score &  &  & \\
	 & sncosmo\_chisq &  &  & \\
	 & sncosmo\_chi2pdof &  &  & \\
	 & sncosmo\_x1 &  &  & \\
	 & sncosmo\_c &  &  & \\
	 \hline
\end{tabular}
\caption{Performance of \tdes for four parameter sets: information only about the host, information available shortly after discovery, information available by the time of peak, and the full parameter set. The performance of \tdes substantially with more data, but high performance is only achieved for the full dataset.}
\label{tab:parameter_subset}
\end{table*}

\bibliography{main}{}
\bibliographystyle{aasjournal}



\end{document}